\documentstyle[12pt,aaspp4]{article}
\lefthead{Kuchner, Reach and Brown}
\righthead{Resonant Structures in the Zodiacal Cloud with DIRBE}
\begin{document}

\title{A Search for Resonant Structures in the Zodiacal Cloud with COBE DIRBE: The Mars Wake and Jupiter's Trojan Clouds}
\author{Marc J. Kuchner}
\affil{Palomar Observatory, California Institute of Technology, Pasadena, CA 91125}
\authoremail{mjk@gps.caltech.edu}
\author{William T. Reach}
\affil{Infrared Processing and Analysis Center, Caltech, Pasadena, CA 91125}
\author{Michael E. Brown\altaffilmark{1}}
\affil{Division of Geological and Planetary Sciences, California Institute of Technology, Pasadena, CA 91125}
\altaffiltext{1}{Alfred P. Sloan Research Fellow}
\author{Running head: Resonant Structures in the Zodiacal Cloud with DIRBE}

\begin{abstract}
We searched the COBE DIRBE Sky and Zodi Atlas for a wake of dust
trailing Mars and for Trojan dust near Jupiter's L5 Lagrange point.  
We compare the DIRBE images to a model Mars wake
based on the empirical model of the Earth's wake as seen by the DIRBE
and place a 3-$\sigma$ upper limit on the fractional overdensity of particles
in the Mars wake of 18\% of the fractional overdensity trailing the Earth.  We place a 3-$\sigma$ upper limit on the effective emitting area
of large (10-100 micron diameter) particles trapped at Jupiter's L5 Lagrange
point of $6 \times 10^{17}$ cm${}^2$, assuming that these large dust grains are distributed in space like the Trojan asteroids.  We would have detected
the Mars wake if the surface area of dust in the wake scaled simply as
the mass of the planet times the Poynting-Robertson time scale.  
\end{abstract}


\section {Introduction}
A planet interacting with a circumstellar dust cloud can produce a variety
of dynamical structures in the dust.  Planets can clear central holes and
create large-scale asymmetries, such as arcs and warps
\markcite{roqu94}(Roques et al. 1994).  Planets can also detain dust in
mean motion resonances, forming structures such as the circumsolar ring
and wake of dust trailing the Earth in its orbit \markcite{jack89}(Jackson
and Zook 1989; \markcite{derm94}Dermott et al. 1994) and clouds of dust at
the planet's Lagrange points \markcite{liou95}(Liou and Zook 1995). 

Understanding the structures of circumstellar debris disks is vital
to the search for extra-solar analogs of our solar system.  Concentrations
in circumstellar dust clouds may confuse planet-finding interferometers
like the Keck Interferometer or the proposed Terrestrial Planet Finder \markcite{beic99}(Beichman 1999).  Smooth exo-zodiacal clouds can be
identified by their symmetry and subtracted from the signal of a Bracewell
interferometer \markcite{macp79}(MacPhie and Bracewell 1979), but cloud
asymmetries can be difficult to distinguish from planets \markcite{beic98}(Beichman 1998).  On the other hand, planet-induced
asymmetries can serve to reveal the presence of a planet that is
otherwise undetectable \markcite{wyat99}(Wyatt et al. 1999).

If we understand the inhomogeneities in our own zodiacal dust cloud, we
will be better prepared to interpret observations of other planetary
systems.  The Diffuse Infrared Background Experiment (DIRBE) aboard the
Cosmic Background Explorer (COBE) satellite has provided detailed,
revealing images of the zodiacal cloud \markcite{spie95}(eg. Spiesman et
al. 1995, \markcite{reac97}Reach et al. 1997).  It surveyed the entire
sky from near-Earth orbit in 10 broad infrared bands simultaneously with
a $0.7^{\circ}$ by $0.7^{\circ}$ field of view over a period of 41 weeks
\markcite{bogg92}(Boggess et al. 1992), and imaged the Earth's ring and wake
\markcite{reac95}(Reach et al. 1995).  We investigated the COBE DIRBE
data set as a source of information about structure in the solar zodiacal
cloud associated with planets other than Earth. 

\section{The Data Set}

We worked with a version of the DIRBE data set which contains the sky
brightness with a model for the background zodiacal emission subtracted:
the zodi-subtracted weekly data set from the DIRBE Sky and Zodi Atlas
(DSZA).  The DIRBE team created the DZSA by fitting an 88-parameter model
of the zodiacal dust emission to the observed sky brightness \markcite{kels98}(Kelsall et al. 1998).  The model includes a smooth
widened-fan component, three pairs of dust bands near the ecliptic plane,
and the Earth's ring and trailing wake, but no structures associated with
Mars, Jupiter, or any other planets.  The zodi-subtracted weekly data
set contains 41 files for each DIRBE band, spanning a period from
10~December 1989 to 21~September 1990, covering 10 bands, centered
at 1.25, 2.2, 3.5, 4.9, 12, 25, 60, 100, 140, and 240 microns.

Galactic emission dominates the zodi-subtracted maps in the mid and
far-infrared near the galactic plane.  Near the ecliptic plane, the
zodi-subtracted maps are dominated by residuals from the subtraction of
the dust bands that are associated with prominent asteroid families
\markcite{reac97}\markcite{kels98} (Reach et al. 1997; Kelsall et al.
1998).  The presence of these bands makes searching for smooth, faint
heliocentric rings of dust near the ecliptic plane impossible.  However,
we could hope to distinguish a blob of dust following a planet across
the sky from other cloud components and from the galactic background by
the apparent motion of the blob during the COBE mission.

The COBE satellite orbited the Earth near the day/night terminator and
repeatedly mapped a swath of the sky extending about 30 degrees before
and behind the terminator (see the COBE DIRBE Explanatory Supplement
\markcite{cobe}(1997) for details).  Each weekly map contains a robust
average of all the week's data and covers a region a little larger than
the daily viewing swath.  This weekly averaging tends to
exclude transient events that would contaminate our final maps, but
should not otherwise significantly affect a search for large features
that move only a few degrees per week.  Figure 1 is a schematic view of
the solar system during week 34 of the mission (9--16 July 1990) showing
the positions of Earth, Mars and Jupiter, and the DIRBE viewing swath
for that week. 

Because DIRBE never imaged the sky within $60^{\circ}$ of the sun, the
orbits of Mercury and Venus, for instance, do not appear in the data.
Mars appeared in the DIRBE viewing swath for 25 weeks of the mission, and
moved $111^{\circ}$ in ecliptic longitude during those weeks.  Jupiter
moved only $40^{\circ}$ in ecliptic longitude during the entire mission,
but this is sufficient to allow some crude background subtraction.  More
distant planets moved less.  Based on these constraints, we decided to
search the weekly maps for dust features following the orbital paths
of Mars and Jupiter.  Figure 2 shows the intersection of the DIRBE viewing
swath with the ecliptic plane throughout the 41 weeks of the mission,
and the ecliptic longitudes of Mars, Jupiter and the Sun during those
weeks.   

\section{The Mars Wake}

A ring of zodiacal dust particles detained in near-Earth resonances
follows the Earth around the sun \markcite{jack89}(Jackson and Zook 1989; \markcite{derm94}Dermott et al. 1994).  This ring consists mainly of
dust in mean-motion resonances where the particles orbit the sun $j$
times every $j+1$ Earth years (j is a whole number).  Smaller trapped
particles experience greater Poynting-Robertson acceleration, so the
equilibrium locations of their orbital pericenters shift closer to
the Earth on the trailing side, where the component of Earth's gravity
that opposes Poynting-Robertson drag is stronger.  The result, averaged
over many particles, appears as a density enhancement in the ring
behind the Earth---a trailing dust wake.  The Earth's wake was detected by
by IRAS \markcite{derm88}\markcite{reac91}(Dermott et al. 1988,
Reach 1991), and later, by DIRBE as an asymmetry in the near-Earth dust
brightness of $\sim 1.1$ MJy ster${}^{-1}$ at 12 microns and $\sim 1.7$
MJy ster${}^{-1}$ at 25 microns \markcite{reac95} (Reach et al. 1995).
We searched the DIRBE data set for a similar wake of dust trailing Mars.

Blackbody dust at the heliocentric distance of Mars has a typical
temperature of $\sim 220$ K; it emits most strongly in the 12 and 25
micron DIRBE bands.  We restricted our exploration to data from these
two bands.  We began by assembling a composite map of the emission
from beyond the solar system, mostly due to stars and dust in the
Galactic plane, by averaging together all the zodi-subtracted weekly
maps in their native COBE quadrilateralized spherical cube coordinates,
a coordinate system that is stationary on the celestial sphere.  We
subtracted this composite map from each of the zodi-subtracted weekly
maps, effectively removing most of the galactic emission and any other
stationary emission except within a few degrees of the galactic plane,
where the emission is so high that detector and pointing instabilities
make our linear subtraction method ineffective.  

The remaining maps, with outlying data removed, had surface brightness
residuals in the range of -1.7 to +1.0 MJy ster${}^{-1}$ at 12 microns,
and -1.6 to 2.1 MJy ster${}^{-1}$ at 25 microns.  For comparison, the
typical total zodiacal background near Mars during the mission is
$\sim 35$ MJy ster${}^{-1}$ at 12 microns and $\sim 66$ MJy ster${}^{-1}$
at 25 microns.  The most prominent remaining features were the stripes
parallel to the ecliptic plane within a few degrees of the ecliptic
plane that are associated with the asteroidal dust bands.  The
next most prominent remaining features were wide bands extending
$\pm 30$ degrees from the ecliptic that appeared to follow the sun.
The $12 - 25$ micron color temperature of the wide bands was
$\sim 280$ K; they are probably residuals resulting from imperfect
subtraction of the Earth's ring and wake.  We assembled a crude map of
the residual near-Earth flux by averaging together the galaxy-subtracted
maps in geocentric ecliptic coordinates referenced to the position of
the Sun.  Subtracting this from the weekly maps cancelled most of
the signal in the wide bands.   Mars moved $87^{\circ}$ with respect
to the Sun during the mission, allowing us to subtract this composite
map without subtracting a significant flux from a wake moving with Mars.
Figure 3a shows our map of the galactic background; Figure 3b shows the
near-Earth residuals.

Next we chose subframes of each weekly map centered on the ecliptic
coordinates of Mars in the middle of the week, and inspected them
visually.  No structure in the data appeared to move with Mars from
week to week.

In order to understand the data better, we constructed a simple model
of the Mars wake from the empirical model of the Earth's trailing wake
fit to the DIRBE data by \markcite{kels98}Kelsall et al. (1998).  The model
has the following form:
\begin{equation}
n = n_0 \exp{\left [ -{{(r-r_{0})^2} \over { 2 \sigma_r^2}}-{{|z|} \over {\sigma_z}} - {{(\theta  - \theta_0)^2} \over {2 \sigma_{\theta}^2}}\right ]}
\end{equation}
where $n$ is the local average of particle number density times particle
cross section, and $r$,$z$, and $\theta$ are cylindrical coordinates in
the plane of the orbit of Mars centered on the sun.  Mars is located at
$r=r_{0}$, $z=0$, $\theta = 0$.  The parameters of the model, $\theta_0$,
$\sigma_r$, $\sigma_z$, $\sigma_{\theta}$, and $n_0$, are the same as
the corresponding parameters for the Earth's wake:
$\theta_0 = -10^{\circ}$, $\sigma_r =0.10$ AU, $\sigma_z=0.091$ AU, $\sigma_{\theta} = 12.1^{\circ}$.  The shaded area trailing Mars in
Figure 1 shows how this model would appear viewed from above the
ecliptic plane.  The \markcite{kels98}Kelsall et al. (1998) Earth wake
has $n_0 = 1.9 \times 10^{-8} \ {\rm AU}^{-1}$, but we chose
$n_0 = 1.08 \times 10^{-8} \ {\rm AU}^{-1}$ so that the density of
the model would be proportional to the local background dust density at
the orbit of Mars.   The model represents what the Earth wake would
look like if it were trailing Mars instead of Earth.

We evaluated the model's surface brightness by computing the line-of-sight integral
\begin{equation}
I_{\lambda} = E_{\lambda} \int n (r,z,\theta) B_{\lambda}(T) ds
\end{equation}
where $B_{\lambda}(T)$ is the Planck function and $E_{\lambda}$ is an
emissivity modification factor prescribed by the COBE model to account
for the deviation of the Earth wake's spectrum from a blackbody;
$E_{12 \mu {\rm m}} = 1.06$, $E_{25 \mu {\rm m}} = 1.00$.  The
temperature of the dust varies with heliocentric distance, $R$, as
$T = 286 \ {\rm K}\ R^{-0.467}$, following the DIRBE model. This
expression is similar to what you would expect for grey-body dust
($T = 278 \ {\rm K}\ R^{-0.5}$). 

In Figure 4, we compare a synthesized image of the model wake with a background-subtracted image of the infrared sky around Mars.  The image
shows the flux in the 25 micron band averaged over weeks 26--34
(14 May 1990 to 15 July 1990) in ecliptic coordinates referenced to
the position of Mars.   The subset of the weekly 25-micron data used in
this image is indicated by the horizontal stripes labeled ``M'' in
Figure 2.  Mars moved 40 degrees in ecliptic longitude
over this period.  The DIRBE team blanked the data within a square about
$2.5^{\circ}$ on a side centered on Mars, and within a $1.5^{\circ}$
radius circle centered on Jupiter.  A software mask in Figure 4
covers the region around Mars affected by this processing. 
This whole region, up to $40^{\circ}$ behind Mars, shows no sign of
a brightness enhancement that we would associate with a wake
of dust trailing Mars.
 
Maps from the later weeks suffer from an oversubtraction due to
imperfections in the \markcite{kels98}Kelsall et al. (1998) zodi
model, visible as the dark region to the lower right.  
Weeks earlier in the mission suffer from a similar undersubtraction.
These artifacts, our primary source of noise, appear to arise
from dust bands at latitudes of $\sim \pm10^{\circ}$, where bands
associated with the Eos asteroids are prominent in the raw DIRBE
data \markcite{reac97}(Reach et al. 1997).  
As the dust from the asteroid belt spirals towards the sun, perturbations
from planets deform the bands.  The Kelsall et al. model includes a
simple model of this dust band which could not take these perturbations
into account.   We chose the span of weeks used to create Figure 4
to minimize these artifacts, which are easily discernible by their
extent in latitude and longitude.

To better compare the model with the data, we focused on a narrow
strip with a height of $3^{\circ}$ in ecliptic latitude, extending from
$8^{\circ}$ ahead of Mars to $39^{\circ}$ behind Mars in ecliptic
longitude.  This strip contains most of the flux in the model wake.  We
averaged together maps from weeks 26--34 prepared as described above
to produce an image of this strip.  In Figure 5, we plot a cut through
this strip, and we compare it with the model, processed in the same
manner as the data.  The data are dominated by residuals from the
ecliptic bands and the Earth's ring, smeared out in the ecliptic plane
by the orbital motion of Mars.  The standard deviation of the data is
0.54 MJy ster${}^{-1}$; although the distribution of the residuals
is not Gaussian, based on this comparison we can place a rough
3-$\sigma$ upper limit on the central peak of the Mars wake of 18\% of
the flux expected from our simple model.

The empirical model of the Earth's wake we have used for comparison to
the Mars dust environment is not an ideal model for the Mars wake.  It
may not even be a good representation of the Earth's wake.  Since COBE
viewed the Earth wake from near the Earth only, the observations constrain
the product $n_0 \sigma_{\theta}$ for the Earth wake, but do not provide
good constraints on either of these parameters alone.  Kelsall et al. \markcite{kels98}(1998) quote a formal error of 28\% on the
determination of $\sigma_{\theta}$.  Calculations for 12 micron particles
suggest that $\sigma_{\theta}$ for the Earth wake might be 40\% lower
than the Kelsall et al. \markcite{kels98}(1988) number; this figure is
based on Figure 5 in Dermott et al. \markcite{derm94}(1994).  Since we
are sensitive to the wake's surface brightness peak as seen from the
Earth, not Mars, using a more compact wake model affects our upper
limits.  Holding $n_0 \sigma_{\theta}$ constant and decreasing
$\sigma_{\theta}$ by 40\% translates into a decrease of our upper limit
to 11\% of one Earth wake.

Mars has 11\% of the mass of the Earth, so we expect it to trap less
dust than the Earth, but not simply 11\% as much dust.  In fact,
there is no simple scaling law that describes how the density of a 
dust ring relates to the size of the planet that traps it.  The density
of the Mars ring is proportional to the capture probability
times the trapping time for each resonance summed over all relevant
resonances and the distribution of particle sizes.  In the adiabatic
theory for resonant capture due to Poynting-Robertson drag, the capture probabilities depend on the mass of the planet compared to the
mass of the star and on the eccentricity of the particle near
resonance and  \markcite{beau94}(Beauge and Ferraz-Mello 1994).  So one
complicating factor is that the orbits of the dust particles are
slightly more eccentric when they pass Mars than when they pass the
Earth; a particle released on the orbit of a typical asteroid, at
2.7 AU with an eccentricity of 0.14, will have an eccentricity of 0.07
as it passes Mars, and an eccentricity of 0.04 when it passes the Earth
\markcite{wyat50}(Wyatt \& Whipple 1950).  The higher eccentricity
makes them harder to trap.

The trapping time scale is proportional to the time it takes for the
resonant interaction to significantly affect the eccentricity and
libration amplitude of the particle.  When the planet has a circular
orbit, these time scales are on the order of the local
Poynting-Robertson decay time \markcite{liou97}(Liou and Zook 1997),
which scales as $r_0^2/\beta$, where $\beta$ is the ratio
of the Sun's radiation-pressure force on a particle to the Sun's
gravitational force on the particle.  Compared to the P-R drag at the
heliocentric distance of the Earth, the Poynting-Robertson drag force
at the orbit of Mars is less for a given particle by a factor of
$1.52^2 = 2.31$.  The small mass of Mars and the higher eccentricities
of the orbits of the incoming particles work against the formation of
a dense ring, but the greater heliocentric distance of Mars compared
to the Earth works in favor of the formation of the ring.

So far our discussion has assumed that the trapping is adiabatic---
that the orbital elements of the particles change on time scales much
longer than the orbital period.  This approximation may not be as good
for trapping by Mars as it is for trapping by the Earth.  Mars has
a greater orbital eccentricity ($e=0.093$) than the Earth ($e=0.017$).
This increases the widths of the zones of resonance overlap, and makes
a larger fraction of dust orbits chaotic \markcite{murr97}(Murray and
Holman 1997).    

Predicting the density of the Mars wake is another step more complex
than predicting the density of the Mars ring.
Compared to the Earth wake, the Mars wake may form closer to the planet
and have a smaller $\sigma_{\theta}$.  Since Mars is less massive than
the Earth, a given particle would need to have a closer interaction
with Mars than with the Earth to receive an impulse from the planet's
gravity that would balance the Poynting-Robertson drag on the particle
(Weidenschilling and Jackson \markcite{weid93}1993).  For this reason,
we expect the trapped particles which form the Mars wake to prefer
resonant orbits with higher $j$ and lower $\phi$ than similar particles
trapped by the Earth, where $\phi$ is the angle between the perihelion
of the orbit of a particle and the longitude of conjunction of the
particle and the planet.  Our upper limit shows that the Mars wake is
less dense than the Earth wake by more than the simple factor of the
mass ratio times the square of the ratio of the semimajor axes
$= 0.11 \times 2.31 = 0.25$.  However, a thorough numerical simulation
which includes the effects we mentioned and others such as resonant
interactions with Jupiter may be the only good way to relate our upper
limit to the dynamical properties of the dust near Mars.  

\section{Trojan Dust}

While the Earth and Mars can collect abundant low eccentricity particles
from all different orbital phases spiraling in from the asteroid belt,
Jupiter orbits in a distinctly different dust environment.  Outside the
asteroid belt, the dust background probably consists mainly of small
particles with high orbital eccentricities: submicron particles released
by asteroids or comets that are kicked by radiation pressure into more
eccentric orbits than their parent bodies (\markcite{berg73}Berg and
Gr\"un 1973; \markcite{mann95}Mann and Gr\"un 1995).  There is also a
stream of submicron particles from the interstellar medium
\markcite{grun94}\markcite{grog96}(Gr\"un et al. 1994, Grogan et al. 1996)
and there are probably a few particles near Jupiter that originated in
the Kuiper belt \markcite{liou96}(Liou et al. 1996).  Jupiter
probably traps many of the small particles in 1:1 mean motion resonances
(Liou and Zook 1995\markcite{liou95}).  However these small trapped
particles should occupy both ``tadpole'' and ``horseshoe''
orbits, without a strong preference for either, and the locations of
their Lagrange points vary with $\beta$ \markcite{murr94}(Murray 1994).
They probably form large, diffuse ring-like clouds which are difficult
for us to detect.

But there is another potential source of dust that could form concentrated
clouds we could hope to detect against the asteroid bands in the DIRBE
data: the Trojan asteroids.  This population of asteroids orbits the Sun
at $\sim 5.2$ AU in 1:1 resonances with Jupiter, librating about Jupiter's
L4 and L5 Lagrange points, roughly $60^{\circ}$ before and behind the planet. 
They number about as many as the main-belt asteroids.

Marzari et al. \markcite{marz97}(1997) have simulated the collisional
evolution of the Trojan asteroids, and concluded that collisions in
the L4 swarm produce on the order of 2000 fragments in the 1--40 km
diameter range every million years.  If we simplisticly assume a
equilibrium size distribution for the produced particles,
$dn \propto a^{-3.5} da$, where a is the particle radius (Dohnanyi 1969),
we find that there are roughly $10^{23}$ particles in the 10--100 micron diameter size range produced every million years.  These large particles
are likely to stay in roughly the same orbits as their parent bodies,
trapped by Jupiter in ``tadpole'' orbits---orbits that librate
around a single Lagrange point.  They could conceivably form detectable
clouds at L4 and L5. 

Liou and Zook (1995) calculated that 2-micron diameter particles will
stay trapped in 1:1 resonances for $\sim5000$ years.  A 20-micron diameter particle at Jupiter's orbit experiences $1/10$ of the Poynting-Robertson acceleration of 2 micron
particles, and will typically stay trapped for 10 times as long
\markcite{schu80}(Schuerman 1980).  Assuming a trapping time of
$5000$ years $\times$ the dust grain diameter/2 microns, and emissivity appropriate for amorphous icy grains
\markcite{back93}(eg. Backman and Paresce 1993), the 10--100
micron diameter particles in the Trojan cloud will emit a total flux, as viewed
from the Earth, of $\sim 3 \times 10^{-4}$ MJy at 60 microns, a few orders
of magnitude below our detection limit.

However, this is a drastic extrapolation and probably a poor guess at
the actual cloud brightness; the size-frequency distribution of the
Trojan asteroids is not well known and dust cloud is probably not near
collisional equilibrium.  Moreover, the total amount of trapped dust
is subject to severe transients, such as the events that produced the
dust bands associated with main belt asteroid families
\markcite{syke86}(Sykes and Greenberg 1986).  For example, a 20-km
diameter Trojan asteroid ground entirely into 10-micron diameter
dust corresponds to a transient cloud  which, as viewed from the Earth,
would produce a 60 micron flux of $\sim 6$ MJy.  A similarly enhanced
cloud might be visible a few percent of the time.

Unfortunately, Jupiter's Lagrange points do not move far with respect to the
galactic background during the COBE mission; L4 moves 10 degrees and L5
moves 50 degrees, as shown in Figure 2.  Only L5, the trailing Lagrange
point, moves far enough during the mission to make subtracting the
galactic background feasible.  There are about half as many L5 Trojans
known as L4 Trojans, but this is probably because the L5 region has
been searched less intensely than the L4 region, not because the L4
and L5 populations are significantly different
\markcite{shoe89}(Shoemaker et al. 1989).

To make a background-subtracted image of the L5 region, we chose two
subsets from the zodi-subtracted data set.  The first, subset A, is
from the beginning of the mission (weeks 5-10) when L5 was in the
viewing swath and approximately stationary on the sky.  The second,
subset B, is the same region of sky, but contains data from half a
year later in the mission (weeks 33--38), when L5 has moved 45 degrees
away, out of the viewing swath.  The average distance from Earth to
L5 is approximately the same during each time period.  These data
sets are depicted in Figure 2.  We focused on data in the 60 micron
band, the band which contains the emission peak for dust at the local
blackbody temperature at 5.2 AU.  To minimize the residuals from
the zodiacal dust model, we used only data from solar elongations
between 65 and 115 degrees, (or between 245 and 295 degrees).
Figure 6 shows an image constructed from data set A and an image
constructed from data set B, and the difference, A$-$B, which is
dominated by residuals from dust bands associated with asteroid bands
and shows no obvious evidence of enhanced emission at L5.  

We made a simple model for a Trojan cloud of large dust particles by
assuming that they occupy the same dynamical space as the Trojan
asteroids themselves, following Sykes \markcite{syke90}(1990).  Sykes
modeled the asteroidal dust bands by showing how particles constrained
to orbits with a given inclination, eccentricity, and semimajor axis
form a cloud when the remaining three orbital elements are randomized.
He then convolved the shapes of these clouds with the distributions of
orbital elements of the asteroids.  In our case, however, the
distribution of orbital elements is much broader and more important in
determining the shape of the final distribution of particles.  Since
there are only about 70 Trojan asteroids whose orbits are well studied,
the distributions of Trojan asteroid orbital parameters have severe
statistical uncertainties.  Therefore we settle for a simple Gaussian
model for the Trojan cloud, using the orbital parameters as a guide to
the parameters of the Gaussian.  In the following calculations, we will
neglect the inclination of Jupiter's orbit relative to the the Earth's
orbit ($1.305^{\circ}$).
 
A typical Trojan asteroid librates around its Lagrange point with a
period of 148 days.   The mean longitude of the asteroid with respect to
Jupiter, $\phi$, oscillates within limits $\phi_{min}$ and $\phi_{max}$,
which can be calculated, according to Yoder et al.
\markcite{yode83}(1983), from 
\begin{equation}
\sin{ \phi_{min} \over 2} = {\sin{( \alpha / 3)} \over B} \qquad \sin{ \phi_{max} \over 2} = {\sin{( \alpha / 3 + 120^{\circ})} \over B}
\end{equation}
where $B= \eta_0 ({{3 \mu} / {2E}})^{1/2}$, $\sin{\alpha}=B$, $\mu = {\rm M}_{\rm Jupiter}/{\rm M}_{\odot} = 0.000955$ and $\eta_0$ = mean motion of Jupiter = 0.01341 rad yr${}^{-1}$. 
These limits are set by $E$, which is a constant of the motion in the
absence of Poynting Robertson drag:
\begin{equation}
E=-{1 \over 6} \Bigl({{d \phi} \over {d t}}\Bigr)^2 - {{\mu \eta_0^2} \over {2 x}} (1+ 4x^3)
\end{equation}
where $ x = |\sin{(\phi/2)}|$. The libration amplitude, $D$, is $\phi_{max} -\phi_{min}$.  We find that the energy constant is approximately
\begin{equation}
E \approx - {{3} \over {2}} \mu \eta_0^2 \bigl(1 - 0.0133 D + 0.2266 D^2 - 0.0392 D^3\bigr)
\end{equation}
for $D \leq 1.3$.

The fraction of time a particle spends at a given phase, or equivalently,
the distribution in phase of an ensemble of particles is given by
\begin{equation}
P_\phi \propto {{1} \over {d \phi / d t}}.
\end{equation}
We can evaluate this as a function of $D$ with the aid of equations (4)
and (5).  For the distribution of dust libration amplitudes, $P_D$, we
used a simple analytic function that approximates the distribution of
libration amplitudes for Trojan asteroids shown in Figure~5 of
Shoemaker et al. \markcite{shoe89}(1989).   When we average $P_{\phi}$
over $P_D$, we find that the L5 dust cloud is distributed in orbital
phase roughly as a Gaussian centered at $\theta_0=59.5^{\circ}$ behind
Jupiter with a dispersion $\sigma_{\theta}=10^{\circ}$. 

The distribution of the dust in heliocentric latitude can be approximated
in a similar way.  If particle in an orbit of given inclination, $i$,
with small eccentricity, spends a fraction of its time, $f$, at latitude,
$\beta$, an ensemble of particles with small eccentricities and evenly
distributed ascending nodes will have a distribution, at a fixed orbital
phase, of
\begin{equation}
P_{\beta} \propto f \propto (\cos^2{\beta} - \cos^2{i})^{-1/2}.
\end{equation}
We takes the inclination distribution of the particles, $P_i$, to be
a simple analytic function that approximates the data for ``independently
discovered Trojans'' shown in Figure 3 of Shoemaker et al. \markcite{shoe89}(1989).
When we average $P_{\beta}$ over $P_i$, we find the distribution in
latitude is roughly a Gaussian with dispersion $\sigma_{\beta}=10^{\circ}$,
and the distribution in height above the ecliptic has a dispersion
$\sigma_z=0.94$ AU.

The radial distribution of Trojans is more complicated to model, since
both librations and epicycles include radial excursions.  The average L5
Trojan eccentricity is 0.063; a particle with this eccentricity orbits at
a range of heliocentric distances, $\Delta r \approx $ 0.66 AU.  In the
course of its librations, a particle with a typical Trojan libration
amplitude, $D=29^{\circ}$, oscillates in semi-major axis over a range
of $\Delta a \approx $0.14 AU.  We are not sensitive to the radial
structure of the Trojan clouds, so we simply model the radial
distribution as a Gaussian with a full width at half maximum of 0.66 AU,
or a dispersion $\sigma_r=0.24$ AU.

Our final model has the form:
\begin{equation}
n = n_0 \exp{\left [ -{{(r-r_{0})^2} \over { 2 \sigma_r^2}}-{{z^2} \over { 2 \sigma_z^2}} - {{(\theta  - \theta_0)^2} \over {2 \sigma_{\theta}^2}}\right ]}
\end{equation}
where $r$, $z$, and $\theta$ are cylindrical coordinates in the plane of the orbit of Jupiter, and the parameters are: $r_0=5.203$ AU, $\sigma_r=0.24$ AU,  $\sigma_z=0.94$ AU, $\theta_0=-59.5^{\circ}$, and $\sigma_{\theta}=9.7^{\circ}$.
We calculated the surface brightness in the same way as we calculated the surface brightness of the model Mars wake, using an emissivity $E_{60 \mu {\rm m}} = 1$ because we are not considering small grains.  The shaded region
at L5 in Figure 1 represents this model as viewed from above the ecliptic plane.
 
In Figure 6, we compare the difference image A$-$B to a synthesized image
of our model cloud.  For this image, $n_0$ is
$3.4 \times 10^{-8} {\rm AU}^{-1}$, corresponding to an effective
emitting surface area at 60 microns of
$3.3 \times 10^{18}$ cm${}^2$, or one 3-km diameter
asteroid ground entirely into 10-micron diameter dust.  Figure 7 compares the
difference image A$-$B and the model image in a different way; it shows
the region within $\pm 10^{\circ}$ of the ecliptic plane averaged in
ecliptic latitude.  The 1-$\sigma$ noise in the data in Figure 7 is
0.09 MJy ster${}^{-1}$.  Based on this, we can place a rough 3-$\sigma$
upper limit on the effective surface area of the large dust grains at L5 of
$\sim 6 \times 10^{17} {\rm cm}^2$. 

\section{Conclusions}

The zodiacal cloud near the ecliptic plane is a complex tapestry of
dynamical phenomena.  We could not detect the Mars wake or Jupiter's
Trojan clouds among the asteroidal dust bands in the DIRBE maps, despite 
the efforts of the DIRBE team to subtract these bands from the maps. 
We would have detected the Mars wake if it had 18\% of the overdensity
of the Earth wake, based on our empirical model for the Earth
wake.  This upper limit illustrates the complexity of relating resonant
structures in circumstellar dust disks to the properties of perturbing
planets.  For instance, we would have detected the Mars wake if the
surface area of the dust in the wake scaled simply with the mass of the
planet times the Poynting-Robertson time scale.  

The Trojan clouds, by our crude estimation, would have been a few orders
of magnitude too faint to detect if the dust concentration in these clouds
were at its mean levels.  However, a transient cloud created by a
recent collision of Trojan asteroids might have been detectable.  We
measured that the total 60-micron flux from large
(10--100 micron diameter) dust particles trapped at Jupiter's
L5 Lagrange point is less than $\sim30$ kJy.

\acknowledgments

We thank  Antonin Bouchez, Eric Gaidos, Peter Goldreich, Renu Malhotra and
Ingrid Mann for helpful discussions, and our referees for their
thoughtful comments.

\newpage

\figcaption{The solar system during week 34.  The shaded regions following Mars represents our model for the Mars wake; the shaded region centered on L5 represents our model for the Trojan cloud.   The hatched area represents the DIRBE viewing swath for that week.  \label{fig1}}  

\figcaption{The ecliptic longitudes of the Sun, Mars, Jupiter, and Jupiter's L4 and L5 Lagrange points during 40 weeks of the COBE mission when DIRBE was recording.  The shaded diagonal stripes show the intersection of the DIRBE viewing swath with the ecliptic plane.  The vertical dashed lines show where the galactic plane crosses the ecliptic. The horizontal bars show the data sets used to construct Figures 4,5,and 6. \label{fig2}}  

\figcaption{Two 25-micron backgrounds that we subtracted from the Mars images. a) The galactic background, constructed by averaging all the weekly maps in their native quadrilateralized spherical cube coordinates.  b) Residuals from the Earth's wake, constructed by averaging all the weekly galaxy-subtracted maps in a geocentric ecliptic coordinate system with the sun at the origin.\label{fig3}}

\figcaption{An image of the sky near Mars at 25 microns, compared to a model based on the COBE DIRBE empirical model for the wake trailing Earth.  The image is averaged over weeks 26--34 (data set M).  The region within $1.5^{\circ}$ of Mars has been covered by a software mask.\label{fig4}}

\figcaption{A cut through the image of the 25-micron sky near Mars shown in Figure 4, compared to the same model.  \label{fig5}}

\figcaption{Raw DSZA images in the ecliptic plane at 60 microns.  L5 is at the center of image A, but it has moved 45 degrees to the right of center in image B.  The difference, A$-$B cancels most of the galactic emission, but is dominated by residuals from dust bands associated with the asteroid belt and does not reveal any Trojan dust.  The model shows what we would expect the difference A$-$B to look like, given some simple assumptions about the Trojan clouds.\label{fig6}}

\figcaption{The difference A$-$B compared to the model for the L5 cloud.  This plot shows a region of the 60-micron maps from Figure 6 within $\pm 10^{\circ}$ of the ecliptic plane that has been averaged in latitude.  Based on this comparison, we place a 3-$\sigma$ upper limit on the surface area of the L5 cloud of $6 \times 10^{17} {\rm cm}^2$.\label{fig7}}

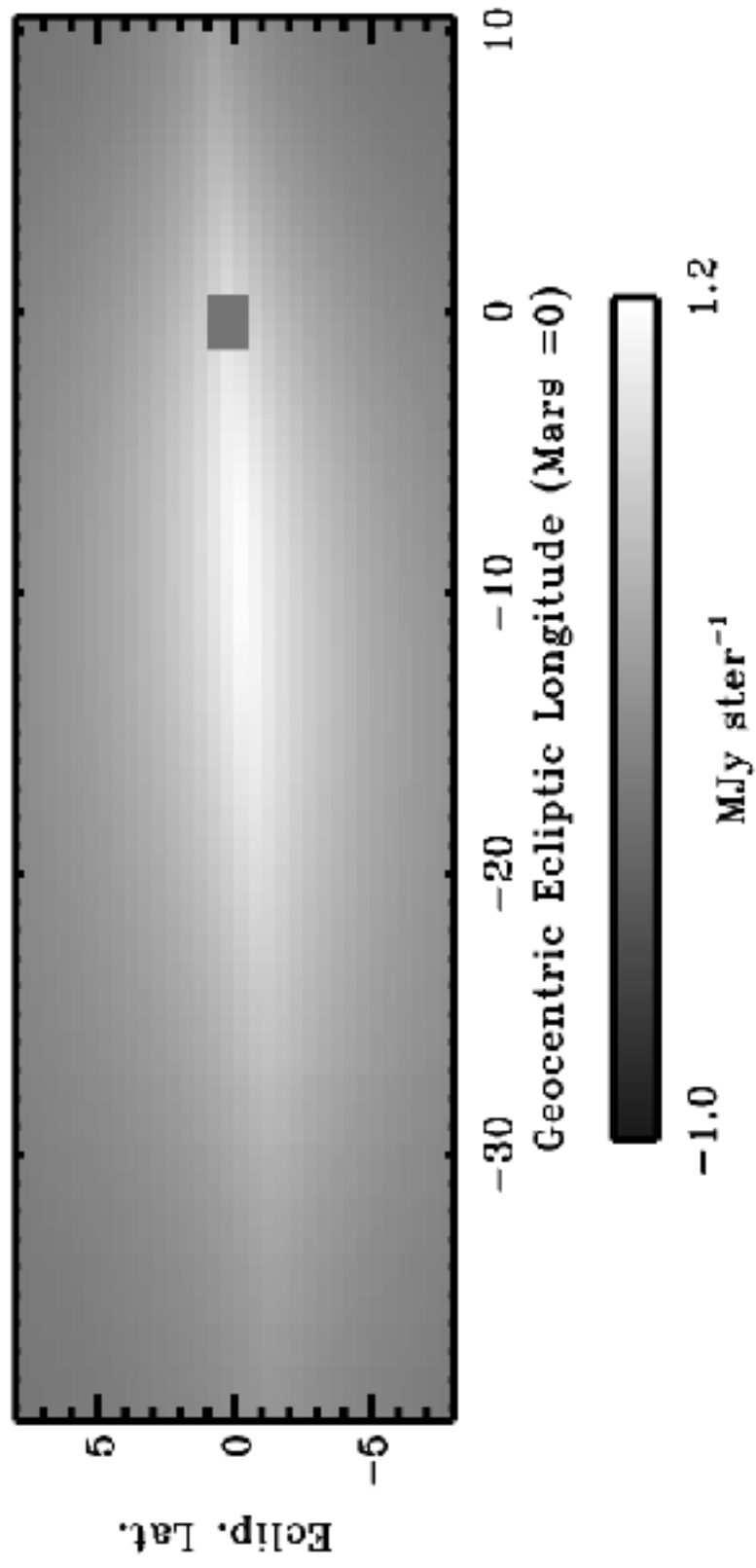
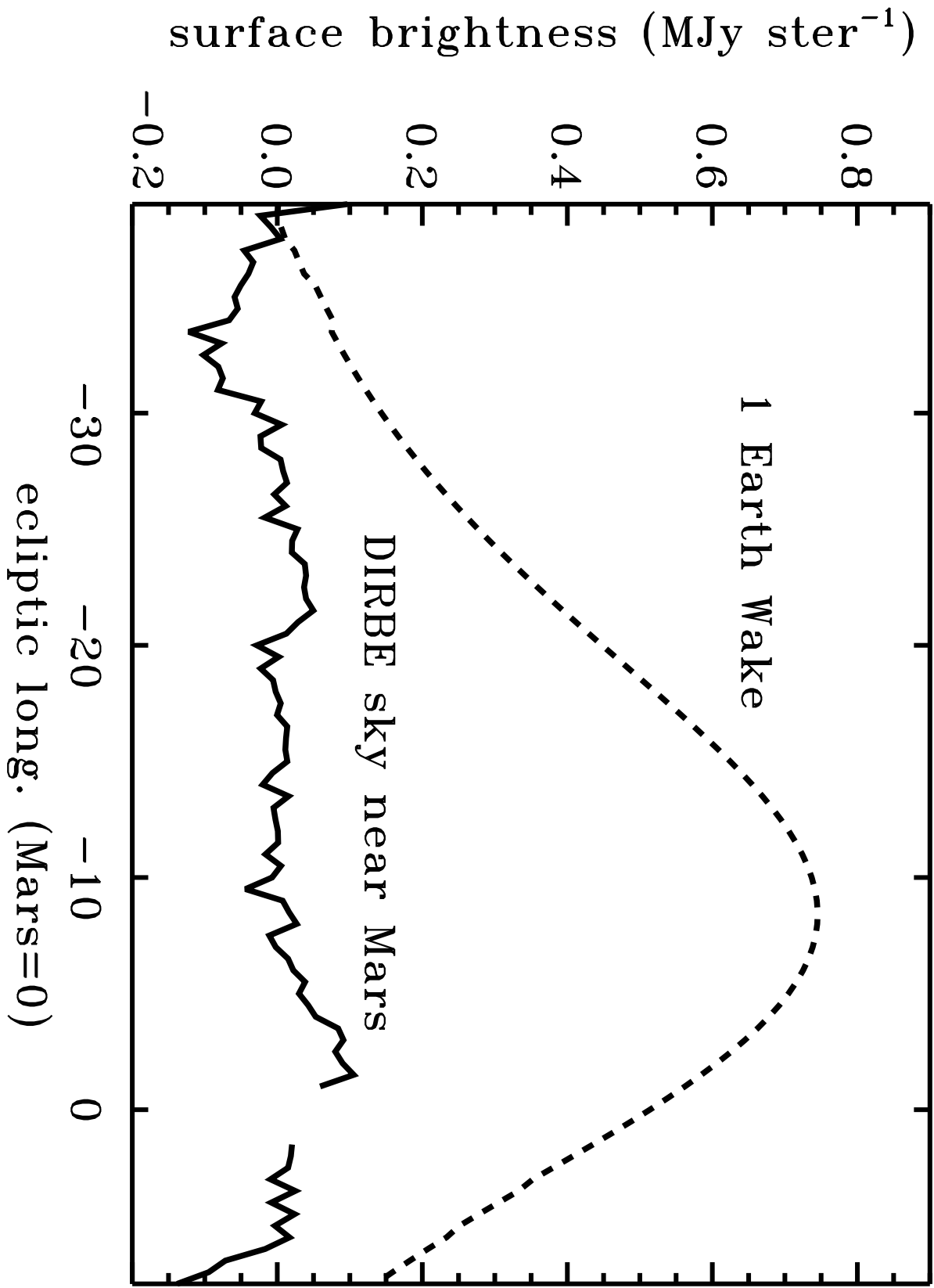
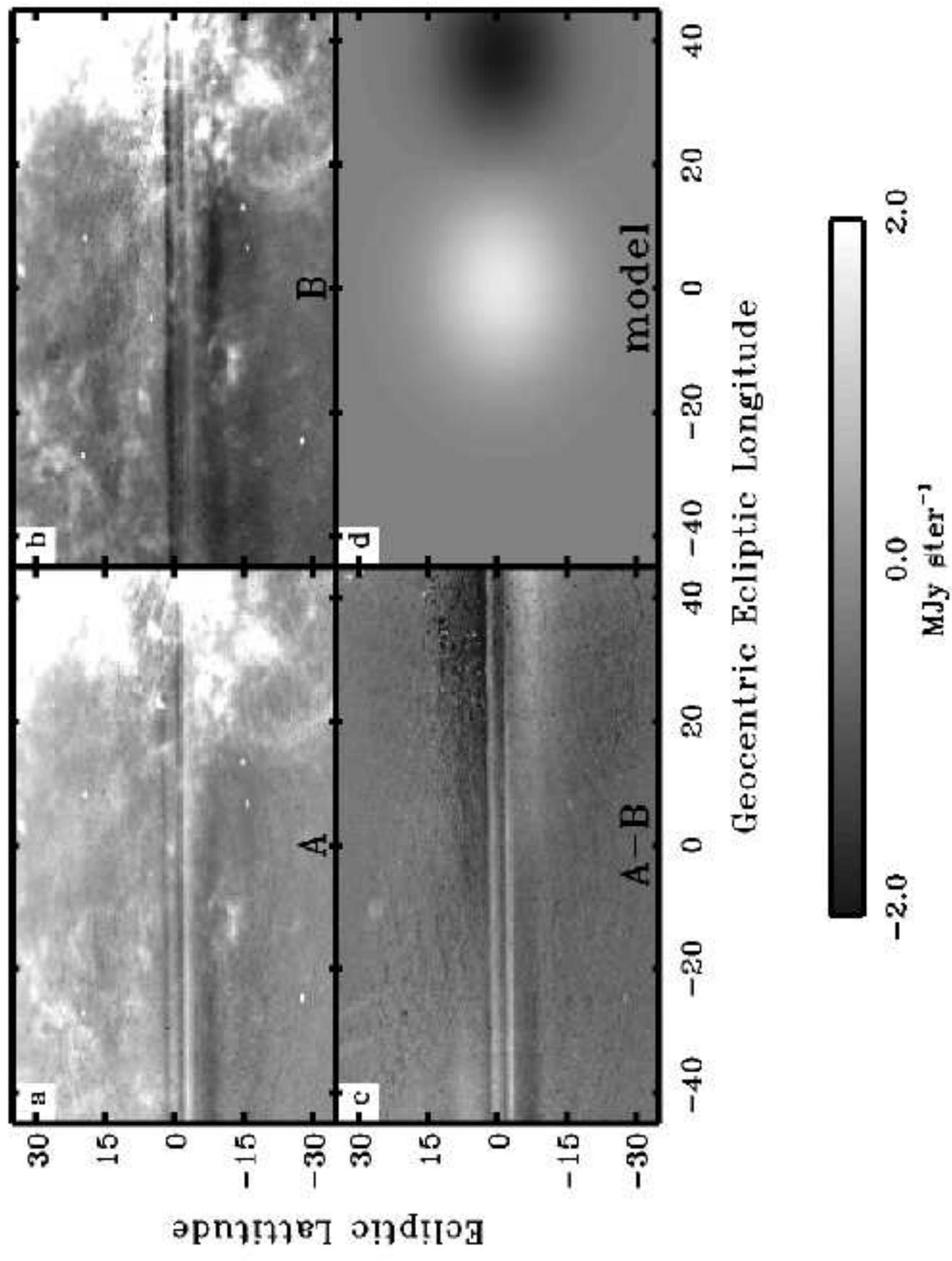
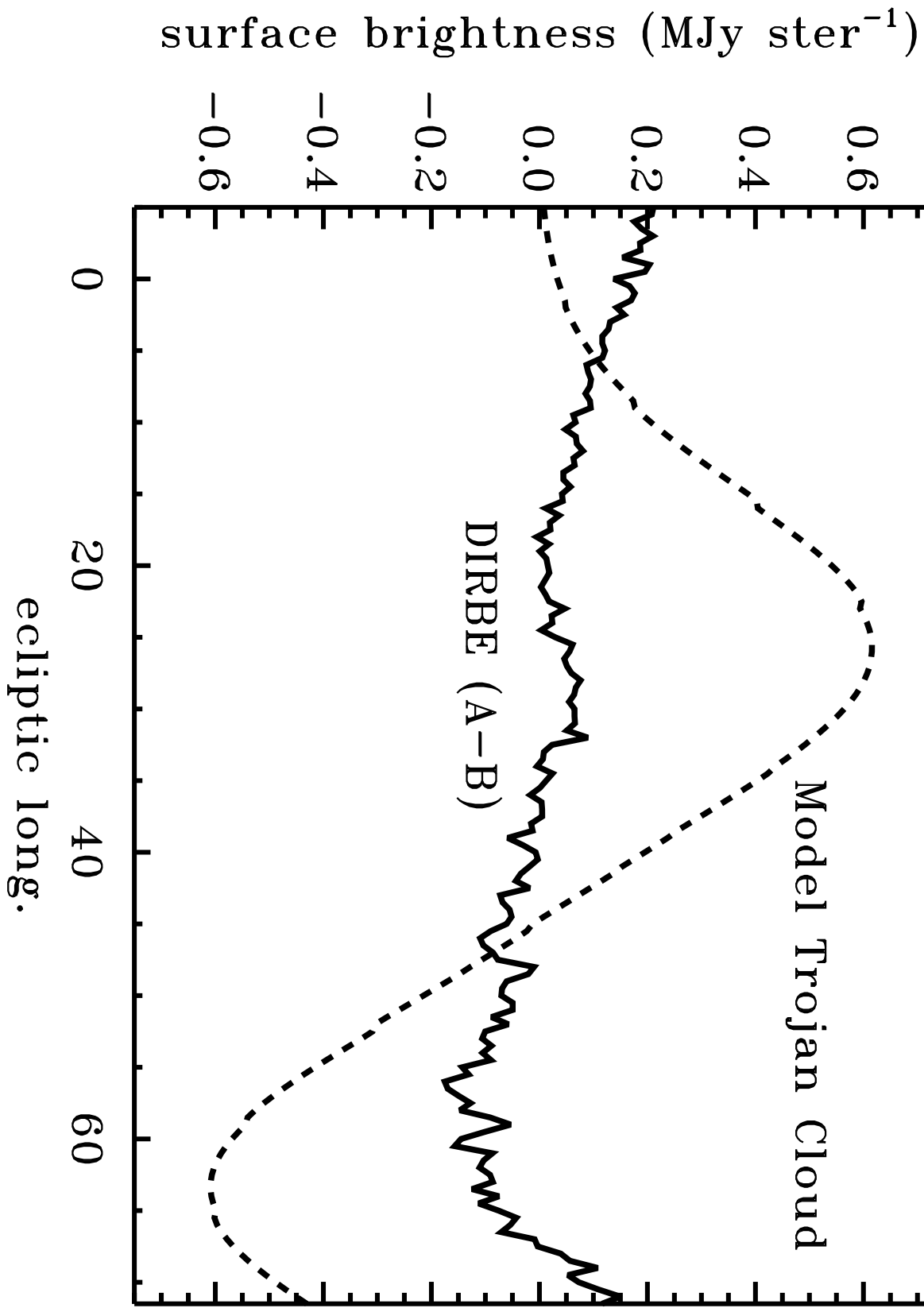
\end{document}